\documentclass[a4paper, 10pt, conference]{ieeeconf}
\usepackage{graphicx}
\usepackage{times}
\usepackage{epsfig}
\usepackage{amsmath}
\usepackage{amssymb}
\usepackage{mathptmx} 

\usepackage{txfonts} 
\usepackage{booktabs}
\usepackage{tabularx}
\usepackage{makecell}
\usepackage{enumerate}
\usepackage{float}
\usepackage{hyperref}

\usepackage{mathtools}
\usepackage[boxed]{algorithm}
\usepackage[noend]{algpseudocode}

\usepackage{listings}
\usepackage{color}

\definecolor{dkgreen}{rgb}{0,0.6,0}
\definecolor{gray}{rgb}{0.5,0.5,0.5}
\definecolor{mauve}{rgb}{0.58,0,0.82}

\usepackage[symbol]{footmisc}

\usepackage{diagbox}

\IEEEoverridecommandlockouts                              

\title{\LARGE \bf RISE-Based Adaptive Control with Mass-Inertia Parameter Estimation for Aerial Transportation of Multi-Rotor UAVs}
\author{Shuyang Shi, Yuzhu Li, Wei Dong}

\begin{document}

\maketitle

\begin{abstract}
This paper proposes an adaptive tracking strategy with mass-inertia estimation for aerial transportation problems of multi-rotor UAVs. The dynamic model of multi-rotor UAVs with disturbances is firstly developed with a linearly parameterized form. Subsequently, a cascade controller with the robust integral of the sign of the error (RISE) terms is applied to smooth the control inputs and address bounded disturbances. Then, adaptive estimation laws for mass-inertia parameters are designed based on a filter operation. Such operation is introduced to extract estimation errors exploited to theoretically guarantee the finite-time (FT) convergence of estimation errors. Finally, simulations are conducted to verify the effectiveness of the designed controller. The results show that the proposed method provides better tracking and estimation performance than traditional adaptive controllers based on sliding mode control algorithms and gradient-based estimation strategies.
\end{abstract}

\section{INTRODUCTION}
In recent years, unmanned aerial vehicles (UAVs), especially multi-rotor UAVs, have been widely applied to military/civil transportation tasks such as parcel delivery \cite{doi:10.1260/1756-8293.7.4.395}, equipment deployment \cite{9665137}, and rescue missions \cite{article_rescue}. These tasks share common characteristics that the mass-inertia parameters of UAVs vary from flight to flight, and a fine-tune of controllers to retain good tracking performances between missions is time-consuming \cite{9268410}. Thus, the varying, namely, the uncertain mass-inertia parameters, can be problematic for the control of UAVs \cite{6220873}. Meanwhile, these light and flexible UAVs are susceptible to external disturbances such as winds \cite{doi:10.1177/0020294019847688}, which seriously degrades flight performances.

To maintain good flight performance of multi-rotor UAVs and better conduct transportation missions, researchers have proposed different control methods to reduce the influence of uncertain mass-inertia parameters and disturbances, such as PID control \cite{doi:10.1007/s10514-012-9280-5}, \cite{doi:10.1142/S2301385014500034}, backstepping control \cite{8657611}, and sliding mode control (SMC) \cite{1234}, just to enumerate a few. These controllers incorporate mass-inertia changes caused by different loads into disturbances \cite{doi:10.1177/16878140211002723} and compensate for the overall disturbances via different methods. For instance, \cite{doi:10.1007/s10514-012-9280-5} studied the stability of UAVs with dynamic load disturbances and improved the control performance by careful selection of control gains. \cite{1234} proposed a non-singular terminal sliding mode controller with high-order sliding mode observers to address uncertain parameters and disturbances. 

While incorporating mass-inertia uncertainties into disturbances is generally effective, improvements are possible. In aerial transportation problems, the variable mass-inertia parameters of loads take a large proportion of the UAV self-parameters, but the ability of the above controllers to accommodate for such variation is moderate \cite{https://doi.org/10.1002/acs.2937}. To tackle such problems, adaptive controllers with mass-inertia estimation have been considered. These controllers provide mass-inertia estimation \cite{7152250} to improve dynamic models while addressing external disturbances. Hence, dynamic models can be corrected after UAVs take off with new loads, yielding better tracking performances \cite{doi:10.1142/S0219878909001977}, \cite{6716436}. For instance, Mellinger {\it et al.} \cite{6094871} proposed an adaptive PID control method that estimates payload parameters during hover via the least-squares methods. In \cite{7839987}, an adaptive cascade controller was designed with the estimation of external force and position of the center of mass. \cite{PALUNKO20112626} established a complete dynamics model of quadrotors and proposed an adaptive controller based on feedback linearization and mass-inertia estimation. Bouadi {\it et al.} \cite{7152250} addressed a SMC algorithm with consideration of white Gaussian noise and mass-inertia uncertainties. \cite{8598841} designed a learning rate-based SMC controller with the estimation of the mass of variable loads for altitude control. \cite{MOFID20181} proposed a finite-time sliding mode controller for disturbance rejection and an adaptive-tuning scheme for mass estimation. In \cite{9341402}, no prior knowledge of uncertain parameters was required via adaptive estimation laws based on signum and saturation functions. \cite{https://doi.org/10.1002/rnc.5760} designed a non-singular fast terminal sliding mode controller based on adaptive integral backstepping to overcome external disturbances and an adaptive estimation algorithm to estimate the variable mass of loads.

However, the methods mentioned can still be improved in two aspects. Firstly, the parameter estimation performance is influenced by disturbances.The convergence of estimation error can not be guaranteed via traditional least-squares \cite{6094871} or gradient-based \cite{PALUNKO20112626} algorithms when external disturbance exists. These methods are sensitive to disturbances and may trigger bursting phenomena, i.e., the estimated parameters may go to infinity, leading to the instability of the system \cite{8978475}. In some improved gradient-based methods, the boundness of estimation error is retained, but the error convergence can not be achieved \cite{7839987}, \cite{7152250}. In \cite{8598841}$\sim$\cite{https://doi.org/10.1002/rnc.5760}, the estimation gradient is determined by high-order tracking errors, which makes the convergence speed easily interfered bu external disturbances. Secondly, the performances of controllers are degraded in practical applications. While SMC has been exploited in \cite{9341402} to address the influence of environmental disturbances, the performance in practical application is not satisfactory enough. The chattering phenomena of sliding mode controllers make the control input signal unreachable physically \cite{8598841}, which significantly degrades the control performances. In \cite{MOFID20181}, \cite{https://doi.org/10.1002/rnc.5760}, such phenomena are restrained via continuous terms in sliding surfaces \cite{https://doi.org/10.1002/rnc.5760}, but the rapid-changing amplitude of control input signals are still hard to achieve in physical systems.

Given the discussion above, this article proposes a new adaptive control method with mass-inertia estimation and disturbance rejection for aerial transportation tasks of multi-rotor UAVs. A RISE term \cite{6579806} is applied for smoothing control inputs of the controller and disturbance rejection. A filter operation \cite{https://doi.org/10.1002/rnc.3247} is introduced to extract estimation errors exploited to guarantee the FT convergence of estimation errors theoretically. Then adaptive estimation laws are designed based on the extracted history estimation errors.

The major contributions of our work are summarized as follows:
\begin{enumerate}[1)]
    \item An adaptive control method based on RISE terms is formulated with mass-inertia estimation. The scheme guarantees the asymptotic convergence of tracking error and FT convergence of estimated parameters under disturbances and provides smooth control input signals achievable in practical applications.
    \item The effectiveness of the proposed method is verified through comparative simulation results with MATLAB.
\end{enumerate}

The rest of this article is organized as follows. A mathematical model of the studied multirotor system is described in Section \uppercase\expandafter{\romannumeral2}. Section \uppercase\expandafter{\romannumeral3} provides the cascade controller design and Section \uppercase\expandafter{\romannumeral4} formulates the parameter update law. Afterward, stability analysis is conducted in Section \uppercase\expandafter{\romannumeral5}. Section \uppercase\expandafter{\romannumeral6} presents comparative simulation results. Finally, conclusions are drawn in Section \uppercase\expandafter{\romannumeral7}.

\section{DYNAMIC MODEL}
To develop the dynamic model of UAV, the defination of frames is first given as is shown in the following figure. The inertial frame (the earth frame) $\{E\}$ is fixed on the ground and the body fixed frame $\{B\}$ is chosen to coincide with the geometric center of the UAV. Let $\eta_1 \triangleq [\begin{matrix}x & y & z \end{matrix}]^T \in \mathbb{R}^3$ denote the position of the origin of $\{B\}$ and $\eta_2 \triangleq [\begin{matrix}\phi & \vartheta & \psi \end{matrix}]^T \in \mathbb{R}^3$ represent the three Euler angles roll, pitch and yaw in frame $\{E\}$. In order to simplify the model, the CoG is assumed to be fixed in $\{B\}$ when loads changes \cite{8598841}. Ignoring the asymmetry of the multi-rotor UAV and according to the Newton-Euler formalism, the rigid body dynamics model used in the subsequent controller design and stability analysis are governed by
\begin{equation}
    \begin{cases}
        m \ddot{\eta}_1 = F - \left [ \begin{matrix} 0 \\ 0 \\ mg\end{matrix} \right] - \Delta_1  \\
        J \ddot{\eta}_2  = \tau_B - \dot{\eta}_2 \times J \dot{\eta}_2 - \Delta_2
    \end{cases}
    \label{equ:dynamics}
\end{equation}
where $m \in \mathbb{R}$ represents the unknown mass of the multirotor; $J \in \mathbb{R}^{3\times3}$ is a matrix representing unknown moment of inertia of the multirotor about the origin of  $\{B\}$. Its non-diagonal elements are set to be zero due to the symmetry of the UAV. $F \in \mathbb{R}^3$ denotes the multirotor force vector expressed in frame $\{E\}$ and $\tau_B \in \mathbb{R}^3$ denotes the torque expressed in frame $\{B\}$. $\Delta_1 \in \mathbb{R}^3$ and $\Delta_2 \in \mathbb{R}^3$ are defined to express the unknown addictive nonlinear disturbances. Equation (\ref{equ:dynamics}) can be partly simplized and rewritten into a more compact form
\begin{equation}
    \begin{cases}
    m \ddot{\eta}_1 + G + \Delta_1 = F \\
    J \ddot{\eta}_2 +C(\dot{\eta}_2)\dot{\eta}_2 +\Delta_2 = \tau_B
    \end{cases}
    \label{equ:rewrite_dynamics}
\end{equation}
by defining $C(\dot{\eta}_2) \triangleq -S(J\dot{\eta}_2)\in \mathbb{R}^{3 \times 3}$ and $G \triangleq[\begin{matrix} 0 & 0 & mg \end{matrix}]^T \in \mathbb{R}^3$, where $S(\cdot) \in \mathbb{R}^{3 \times 3}$ represents the skew symmetric matrix of a vector. The rewritten equation (\ref{equ:rewrite_dynamics}) is still in a seperate form because the following controller and parameter update law design are developed in a cascade manner.

There are several properties and assumptions of the dynamics model which will be exploited in the subsequent development:

\noindent {\bf Property 1.} Part of the dynamics equation (\ref{equ:rewrite_dynamics}) can be linearly parameterized as
\begin{equation}
    \begin{cases}
    \Psi_1 \theta_1 \triangleq m \ddot{\eta}_1 + G\\
    \Psi_2 \theta_2 \triangleq J \ddot{\eta}_2 +C(\dot{\eta}_2)\dot{\eta}_2
    \end{cases}
    \label{equ:linearize}
\end{equation}
where $\theta_1 \in \mathbb{R}$ and $\theta_2 \in \mathbb{R}^{3 \times 3}$ contain the unknown system mass-inertia parameters, $\Psi_1(\ddot{\eta}_1) \in \mathbb{R}^3$ and $\Psi_2(\dot{\eta}_2, \ddot{\eta}_2) \in \mathbb{R}^{3 \times 3}$ are the regression matrices which contains known functions of measured acceleration, angular rate and angular acceleration respectively. The above linearization can also be formulated with desired position and attitude vectors, yielding 
\begin{equation}
    \begin{cases}
    \Psi_{1d} \theta_1 \triangleq m \ddot{\eta}_{1d} + G\\
    \Psi_{2d} \theta_2 \triangleq J \ddot{\eta}_{2d} +C(\dot{\eta}_{2d})\dot{\eta}_{2d}
    \end{cases}
    \label{equ:linearize_d}
\end{equation}
where $\Psi_{1d}(\ddot{\eta}_{1d}) \in \mathbb{R}^3$ and $\Psi_{2d}(\dot{\eta}_{2d}, \ddot{\eta}_{2d}) \in \mathbb{R}^{3 \times 3}$ are bounded desired regression matrices containing known functions of desired tracking vectors respectively.

\noindent {\bf Assumpition 1.} The regression matrices $\Psi_{1d}$ and $\Psi_{2d}$ defined above satisfy the PE condition described in \cite{doi:10.1080/00207178708933715}, which can be easily fulfilled in our experiments. And the condition is important for the parameter update law given later in the article.

\noindent {\bf Assumpition 2.} The nonlinear disturbances $\Delta_1$ and $\Delta_2$ and their first two-order time derivatives, i.e. $\dot{\Delta}_i$, $\ddot{\Delta}_i$, $(i = 1,2)$ are bounded by known constants.

\section{CONTROL DESITN}
The control objective is to design a controller which guarantees that the system tracks a desired trajectory $\eta_{1d}$ and $\psi_d$ despite the bounded disturbances and uncertain parameters in the dynamics model. The desired trajectory $\eta_{1d}$ and $\psi_d$ are designed such that $\eta^{(i)}_{1d}(t)$ and $\psi^{(i)}_d(t)$, $i = 0, 1, ...4$ exist and are bounded.

\begin{figure}[htbp]
    \centering
    \includegraphics[width=\linewidth]{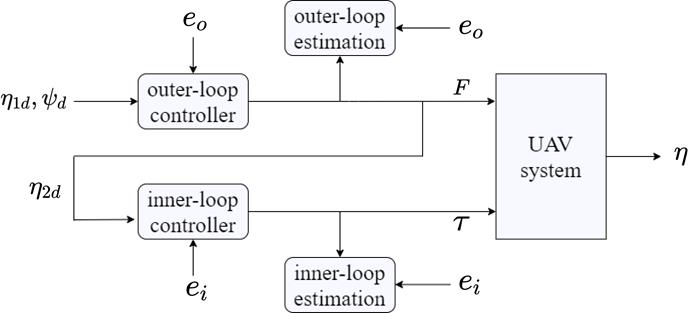}
    \caption{Controller}
    \label{fig:controller}
\end{figure}

The controller illustrated in \ref{fig:controller} is constructed with a cascade structure consisting of an outer-loop controller and an inner-loop controller. The outer-loop controller generates the thrust $F$ and desired roll, pitch angles to track the desired position trajectory and yaw angle. The inner-loop controller is designed to generate the torque $T$ needed to track the desired yaw angle and the calculated roll and pitch angle trajectories.

\subsection{Outer-Loop Controller}
To quantify the control performance, tracking error $e_{o1} \in \mathbb{R}^3$, and two auxiliary filtered tracking errors $e_{o2} \in \mathbb{R}^3$ and $r_{o} \in \mathbb{R}^3$ are defined as follows:
\begin{equation}
    \begin{aligned}
    & e_{o1} \triangleq \eta_{1d} - \eta_1, \\
    & e_{o2} \triangleq \dot{e}_{o1} + k_{o1}e_{o1}, \\
    & r_{o} \triangleq \dot{e}_{o2} + k_{o2}e_{o2}, 
    \end{aligned}
    \label{equ:outer_error}
\end{equation}
where $k_{o1}$, $k_{o2}$ $\in \mathbb{R}^+$ are designed constant control gains. By substituting the errors in (\ref{equ:outer_error}) and the linearized form \ref{equ:linearize_d} into the first dynamic equation in (\ref{equ:rewrite_dynamics}), the open-loop error dynamics of outer-loop system can be developed as:
\begin{equation}
    mr_{o} = \Psi_{1d} \theta_1 + S_1 + \Delta_1 - F
    \label{equ:outer_error_dynamics}
\end{equation}
where the auxiliary function $S_1 \in \mathbb{R}^3$ is defined as
\begin{equation}
    S_1 \triangleq m(k_{o1} \dot{e}_{o1} + k_{o2}\dot{e}_{o2})
    \label{equ:S_1}
\end{equation}
The output force $F$ can be designed with an adaptive feedforward term and a RISE feedback term as
\begin{equation}
    F \triangleq \Psi_{1d} \hat{\theta}_1 + \mu_1
    \label{equ:outer_input}
\end{equation}
In (\ref{equ:outer_input}), $\hat{\theta}_1 \in \mathbb{R}$ denotes the adaptive estimate for the unknown parameter $\theta_1$ whose implementation will be discussed with detail in Section \uppercase\expandafter{\romannumeral4} later. $\mu_1 \in \mathbb{R}^3$ represents the RISE feedback term described in \cite{1310482}, which is designed as
\begin{equation}
    \begin{aligned}
    \mu_1 \triangleq &(k_{s1} +1)e_{o2} - (k_{s1} +1)e_{o2}(0) \\ 
    &+ \int_0^t[(k_{s1} + 1)k_{o2}e_{o2}(\tau) + \beta_o sgn(e_{o2}(\tau))] d\tau
    \end{aligned}
    \label{equ:outer_RISE}
\end{equation}
where $k_{s1} \in \mathbb{R}^+$ and $\beta_o \in \mathbb{R}^+$ are constant control gains and $sgn(\cdot)$ represents the signum function. Then the time derivative of the RISE term can be derived as
\begin{equation}
    \dot{\mu}_1 = (k_{s1} + 1)r_{o} + \beta_o sgn(e_{o2})
    \label{equ:outer_RISE_d}
\end{equation}
Substituting equation (\ref{equ:outer_input}) into (\ref{equ:outer_error_dynamics}), the closed-loop error dynamics of outer-loop system can be developed as 
\begin{equation}
    mr_o = \Psi_{1d} \tilde{\theta}_1 + S_1 + \Delta_1 - \mu_1
    \label{equ:outer_close_error_dynamics}
\end{equation}
where $\tilde{\theta}_1 \triangleq \theta_1 - \hat{\theta}_1 \in \mathbb{R}$ denotes the parameter estimation error.
Equation (\ref{equ:outer_close_error_dynamics}) will be exploited in Section \uppercase\expandafter{\romannumeral5} to facilitate stability analysis of outer-loop controller.

The desired body attitude in frame $\{E\}$ then can be derived in the same manner as \cite{5980409}, by first calculating the desired z-axis direction of the body frame $\{B\}$, which is alone the desired output force $F$:
\begin{equation}
    {\bf z}_B = \frac{F}{||F||}
    \label{equ:zB}
\end{equation}
where $||\cdot||$ denotes the Euclidean norm. $||F||$ will be nonzero to avoid free-falling. Given the desired yaw angle $\psi_d$, a unit vector ${\bf x}_C \in \mathbb{R}^3$ can be defined as
\begin{equation}
    {\bf x}_C \triangleq [\begin{matrix} -s\psi_d & c\psi_d & 0 \end{matrix}]^T
    \label{equ:xc}
\end{equation}
where $s \psi_d$ and $c \psi_d$ denotes $sin(\psi_d)$ and $cos(\psi_d)$ respectively. Provided ${\bf x}_C \times {\bf z}_B \neq 0$, the orientation of frame $\{B\}$ can be uniquely determined as
\begin{equation}
    \begin{aligned}
    & {\bf x}_B = \frac{{\bf x}_C \times {\bf z}_B}{||{\bf x}_C \times {\bf z}_B||} \\
    & {\bf y}_B = {\bf z}_B \times {\bf x}_B \\
    & R_B^E \triangleq [\begin{matrix} {\bf x}_B & {\bf y}_B &{\bf z}_B\end{matrix}]
    \end{aligned}
    \label{equ:frameB}
\end{equation}
where ${\bf x}_B$ and ${\bf y}_B$ are x and y axes of frame $\{B\}$ respectively. $R_B^E \in \mathbb{R}^{3 \times 3}$ is the rotation matrix from the body-fixed frame $\{B\}$ to the inertial frame $\{E\}$, given by
\begin{equation}
    R_B^E = 
    \left[ \begin{matrix}
    c\psi c\vartheta & c\psi s\vartheta s\phi - s\psi c\phi & c\psi s\vartheta c\phi + s\psi s\phi \\
    s\psi c\vartheta & s\psi s\vartheta s\phi + c\psi c\phi & s\psi s\vartheta c\phi - c\psi s\phi \\
    -s\vartheta & c\vartheta s\phi & c\vartheta c\phi
    \end{matrix} \right]
    \label{equ:rot}
\end{equation}
The desired roll angle $\phi_d$ and pitch angle $\theta_d$ can be calculated from $\psi_d$ and $F$ via equations (\ref{equ:zB}), (\ref{equ:xc}), (\ref{equ:frameB}) and (\ref{equ:rot}).
\subsection{Inner-Loop Controller}
The realization of inner loop controller is similar to that of the outer controller. First, tracking error $e_{i1} \in \mathbb{R}^3$ and auxiliary filtered errors $e_{i2}$, $r_{i} \in \mathbb{R}^3$ are defined as
\begin{equation}
    \begin{aligned}
    & e_{i1} \triangleq \eta_{2d} - \eta_2, \\
    & e_{i2} \triangleq \dot{e}_{i1} + k_{i1}e_{i1}, \\
    & r_{i} \triangleq \dot{e}_{i2} + k_{i2}e_{i2}, 
    \end{aligned}
    \label{equ:inner_error}
\end{equation}
where $k_{i1} \in \mathbb{R}$ and $k_{i2} \in \mathbb{R}$ are constant control gains. By substituting the errors in (\ref{equ:inner_error}) and the linearized form \ref{equ:linearize_d} into the second equation in (\ref{equ:rewrite_dynamics}), the open-loop error dynamics of outer-loop system can be developed as:
\begin{equation}
    Jr_{i} = \Psi_{2d} \theta_2 + S_2 + \Delta_2 - \tau_B
    \label{equ:inner_error_dynamics}
\end{equation}
where the auxiliary function $S_1 \in \mathbb{R}^3$ is defined as
\begin{equation}
    \begin{aligned}
    S_2 \triangleq & J(k_{i1} \dot{e}_{i1} + k_{i2}\dot{e}_{i2}) + C(\dot {\eta}_2)\dot \eta_2 - C(\dot {\eta}_{2d})\dot \eta_{2d}
    \end{aligned}
    \label{equ:S_2}
\end{equation}
The inner-loop control output $T$ can be designed with an adaptive feedforward term and a RISE feedback term as
\begin{equation}
    \tau_B \triangleq \Psi_{2d} \hat{\theta}_2 + \mu_2
    \label{equ:inner_input}
\end{equation}
In (\ref{equ:inner_input}), $\hat{\theta}_2 \in \mathbb{R}$ denotes the adaptive estimate for the unknown parameter $\theta_2$; $\mu_2 \in \mathbb{R}^3$ represents the RISE feedback term and is designed similar to equation (\ref{equ:outer_RISE}) as
\begin{equation}
    \begin{aligned}
    \mu_2 \triangleq &(k_{s2} +1)e_{i2} - (k_{s2} +1)e_{i2}(0) \\ 
    &+ \int_0^t[(k_{s2} + 1)k_{i2}e_{i2}(\tau) + \beta_i sgn(e_{i2}(\tau))] d\tau
    \end{aligned}
    \label{equ:inner_RISE}
\end{equation}
and its time derivative similar to (\ref{equ:outer_RISE_d}) as
\begin{equation}
    \dot{\mu}_2 = (k_{s2} + 1)r_{i} + \beta_i sgn(e_{i2})
    \label{equ:inner_RISE_d}
\end{equation}
where $k_{s2} \in \mathbb{R}^+$ and $\beta_i \in \mathbb{R}^+$ are constant control gains. Substituting equation (\ref{equ:inner_input}) into (\ref{equ:inner_error_dynamics}), the closed-loop error dynamics of outer-loop system can be developed as 
\begin{equation}
    Jr_i = \Psi_{2d} \tilde{\theta}_2 + S_2 + \Delta_2 - \mu_2
    \label{equ:inner_close_error_dynamics}
\end{equation}
where $\tilde{\theta}_2 \triangleq \theta_2 - \hat{\theta}_2 \in \mathbb{R}^3$ denotes the parameter estimation error. Equation (\ref{equ:inner_close_error_dynamics}) will be exploited in Section \uppercase\expandafter{\romannumeral5} to facilitate stability analysis of inner-loop controller.

\section{PARAMETER ESTIMATION}
The parameter estimation is conducted in both of the control loops with the same error extraction process. In the outer-loop, $\hat {\theta}_1$ is calculated and exploited to generate control outputs, $\hat {\theta}_2$ in the inner-loop respectively.
\subsection{Estimation In Outer-Loop}
The estimation starts with the defination of two filtered auxiliary vectors $F_f$, $\Psi_{1f} \in \mathbb{R}^3$ as the solutions to the following equation
\begin{equation}
    \begin{cases}
    \begin{aligned}
        &\alpha_1 \dot{F}_f + F_f = F, \quad F_f(0) = \bf0 \\
        &\alpha_1 \dot{\Psi}_{1f} + \Psi_{1f} = \Psi_1, \quad \Psi_{1f} = \bf0
    \end{aligned}
    \end{cases}
    \label{equ:filter_variable_outer}
\end{equation}
where $\alpha_1 \in \mathbb{R}^+$ is a designed constant. Another filtered variable only used for analysis $\Delta_{1f} \in \mathbb{R}^3$ is also defined as
\begin{equation}
    \alpha_1 \dot{\Delta}_{1f} + \Delta_{1f} = \Delta_1, \quad \Delta_{1f} = \bf0
    \label{equ:filter_Delta1}
\end{equation}
where $\Delta_{1f}$ is bounded given that $\Delta_1$ is bounded. (\ref{equ:filter_variable_outer}) and (\ref{equ:filter_Delta1}) acctually exert the same low-pass filter operation on both sides of the linearized dynamic model
\begin{equation}
    \Psi_1 \theta_1 +\Delta_1 = F   
\end{equation}
Then a filtered form of the above equation can be expressed as
\begin{equation}
    \Psi_{1f} \theta_{1} +\Delta_1 = F_f
    \label{equ:PFf}
\end{equation}
To extract the estimation error $\tilde{\theta}_1$, $P_1$, $Q_1 \in \mathbb{R}$ are defined as
\begin{equation}
    \begin{cases}
    \begin{aligned}
    & P_1 \triangleq \int_0^t e^{-l_1(t-\tau)}\Psi_{1f}^T(\tau) \Psi_{1f}(\tau) d \tau + \varrho_1\\
    & Q_1 \triangleq \int_0^t e^{-l_1(t-\tau)}\Psi_{1f}^T(\tau) F_f(\tau) d \tau
    \end{aligned}
    \end{cases}
    \label{equ:PQ_outer}
\end{equation}
which are the solutions to the equation below
\begin{equation}
    \begin{cases}
    \begin{aligned}
    & \dot{P}_1 = -l_1 P_1 + \Psi_{1f}^T \Psi_{1f}, \quad P_1(0) = \varrho_1\\
    & \dot{Q}_1 = -l_1 Q_1 + \Psi_{1f}^T F_f, \quad Q_1(0) = 0
    \end{aligned}
    \end{cases}
    \label{equ:dPQ_outer}
\end{equation}
where $l_1 \in \mathbb{R}^+$ is a designed constant, and $\varrho_1 \in \mathbb{R}^+$ is a positive constant selected to ensure $P_1(0)$ is inversible at time $t = 0$. Such definition of $P_1$ yields the following property: 

\noindent {\bf Property 3.} $P_1$ is a positive variable satisfying $0 < \varrho_1 < P_1$. Then $P_1^{-1}$ is globally invertible provided that the offset value $\varrho_1$ is not selected as $0$. The proof of this property is similar to that of \cite{https://doi.org/10.1002/rnc.3247}.


\noindent Similar to $P_1$ and $Q_1$, $\bar{\Delta}_1 \in \mathbb{R}$ is defined as
\begin{equation}
    \bar{\Delta}_1 \triangleq - \int_0^t e^{-l_1(t-\tau)}\Psi_{1f}^T(\tau) \Delta_1 d \tau + \varrho_1 \theta_1
\end{equation}
which is bounded by $||\bar{\Delta}_1||\leq \xi_{\Delta_1}$, where $\xi_{\Delta_1} \in \mathbb{R}^+$ is a positive constant, since the regression vector $\Psi_{1f}$ is locally bounded and $\Delta_1$ is bounded.
Substituting the linearized form (\ref{equ:linearize}) into system dynamics (\ref{equ:rewrite_dynamics}), and substituting equation (\ref{equ:filter_variable_outer}), (\ref{equ:PFf}) into (\ref{equ:PQ_outer}), yields
\begin{equation}
    Q_1 = P_1 \theta_1 - \bar{\Delta}_1
    \label{equ:Q1-P1}
\end{equation}
Then the estimation error is extracted by defining $H_1 \in \mathbb{R}$ as
\begin{equation}
    H_1 \triangleq P_1 \hat{\theta}_1 - Q_1
    \label{equ:H_1}
\end{equation}
which contains the estimation error $\tilde{\theta}_1$ as 
\begin{equation}
    H_1 = -P_1 \tilde{\theta}_1 + \bar{\Delta}_1
    \label{equ:H1_delta}
\end{equation}
is derived by substituting equation (\ref{equ:Q1-P1}) into (\ref{equ:H_1}).

Based on the extracted estimation error above, the parameter update law can be designed as
\begin{equation}
    \dot{\hat{\theta}}_1 = -\gamma \left( \gamma_1 H_1 +\mathrm{sat}\left(H_1\right)\right)
    \label{equ:update_law_outer}
\end{equation}
where $\gamma$, $\gamma_1 \in \mathbb{R}^+$ are positive learing gains and the saturation function $\mathrm{sat}(\cdot): \mathbb{R} \rightarrow \mathbb{R}$ is defined as
\begin{equation}
    \mathrm{sat}(x) = \left \{
    \begin{matrix}
    \begin{aligned}
    &1, \quad x>1 \\
    &x, \quad |x|\leq 1 \\
    &-1, \quad x<-1 \\
    \end{aligned}
    \end{matrix}
    \right.
    \label{equ:sat}
\end{equation}
\subsection{Estimation In Inner-Loop}
In the same manner as estimation in outer-loop, filtered auxiliary vectors $\Psi_{2f} \in \mathbb{R}^{3\times 3}$, $R_f$ amd $\Delta_{2f} \in \mathbb{R}^3$ are defined by the following differential equations:
\begin{equation}
    \begin{cases}
    \begin{aligned}
    &\alpha_2 \dot{\tau}_{Bf} + \tau_{Bf} = \tau_B, \quad \tau_{Bf}(0) = \bf0 \\
    &\alpha_2 \dot{\Psi}_{2f} + \Psi_{2f} = \Psi_2, \quad \Psi_{2f} = \bf0 \\
    &\alpha_2 \dot{\Delta}_{2f} + \Delta_{2f} = \Delta_2, \quad \Delta_{2f} = \bf0
    \end{aligned}
    \end{cases}
    \label{equ:filter_variable_inner}
\end{equation}
where $\alpha_2 \in \mathbb{R}^+$ is a designed constant. For estimation error extraction, $P_2 \in \mathbb{R}^{3 \times 3}$, $Q_2 \in \mathbb{R}^3$, and$\bar{\Delta}_2 \in \mathbb{R}^3$ are defined as
\begin{equation}
    \begin{cases}
    \begin{aligned}
    & P_2 \triangleq \int_0^t e^{-l_2(t-\tau)}\Psi_{2f}^T(\tau) \Psi_{2f}(\tau) d \tau  + \varrho_2 E_3\\
    & Q_2 \triangleq \int_0^t e^{-l_2(t-\tau)}\Psi_{2f}^T(\tau) \tau_{Bf}(\tau) d \tau \\
    & \bar{\Delta}_2 \triangleq - \int_0^t e^{-l_2(t-\tau)}\Psi_{2f}^T(\tau) \Delta_2 d \tau +\varrho_2 E_3 \theta_2
    \end{aligned}
    \end{cases}
    \label{equ:PQ_inner}
\end{equation}
where $l_2$, $\varrho_2 \in \mathbb{R}^+$ are designed positive constants and $E_3 \in \mathbb{R}^{3\times 3}$ is the identity matrix. $\bar{\Delta}_2$ is bounded by $||\bar{\Delta}_2|| \leq \xi_{{\Delta}_2}$. Simlar to $P_1$, $P_2$ has the has the following property:

\noindent {\bf Property 4.} $P_2$ is a positive definite matrix satisfying $0 < \varrho_2 < \lambda_m(P_2)$ where $\lambda_m(P_2)$ is the minimum eigenvalue of $P_2$. And $P_2^{-1}$ is globally invertible.

\noindent Define $H_2 \in \mathbb{R}^3$ as
\begin{equation}
    H_2 \triangleq P_2 \hat{\theta}_2 - Q_2
    \label{equ:H_2}
\end{equation}
which yields
\begin{equation}
    H_2 = -P_2 \tilde{\theta}_2 + \bar{\Delta}_2
    \label{equ:H2_Delta}
\end{equation}
in the same manner as estimation in outer-loop.
The parameter update law for inner-loop can be designed as 
\begin{equation}
    \dot{\hat{\theta}}_2 = -\Gamma\left(\sigma_1 H_2 +\sigma_1 \frac{P_2^TH_2}{||P_2||} + \sigma_2 \frac{P_2^TH_2}{||P_2||\cdot ||H_2||}\right)
    \label{equ:update_law_inner}
\end{equation}
where $\Gamma \in \mathbb{R}^{3\times3}$ is a positive definite diagonal matrix, and $\sigma_1$, $\sigma_2 \in \mathbb{R}^+$ are positive constants.

\section{STABILITY ANALYSIS}
The stability analysis for the proposed methed is conducted in two parts: outer-loop and inner-loop. Both controllers yields asymptotic convergence of tracking error and finite time convergence of estimation error.
\subsection{Inner-Loop Analysis}
To facilitate stability analysis of inner-loop, the time derivative of equation (\ref{equ:inner_close_error_dynamics}) is exploited:
\begin{equation}
    J\dot{r}_i = \tilde{N}_i +N_{{\Delta}_i} - \dot{\mu}_2 - e_{i2}
    \label{equ:inner_close_error_d}
\end{equation}
In equation (\ref{equ:inner_close_error_d}), part of the equation is seperated into two unmeasurable auxiliary functions $\tilde{N}_i$, $N_{\Delta_i}$ $\in \mathbb{R}^3$ which are upper-bounded by different terms. The motivation for such operation has been discussed in \cite{5586643}. 
Substituting equation (\ref{equ:update_law_inner}) into equation (\ref{equ:inner_close_error_d}), $\tilde{N}_i$ and $N_{\Delta_i}$ can be defined as
\begin{equation}
    \begin{aligned}
    & \tilde{N}_i(t) \triangleq \dot S_2 + e_{i2} + N_i \\
    & N_{\Delta_2} \triangleq \dot \Delta_2
    \end{aligned}
    \label{equ:Ni}
\end{equation}
where $N_i \in \mathbb{R}^3$ is another auxiliary function defined as
\begin{equation}
    N_i \triangleq \dot{\Psi}_{2d} \tilde{\theta}_2 - \Psi_{2d}\dot{\hat{\theta}}_2
\end{equation}
As is discussed in \cite{1310482}, $\tilde N_i$ is upper bounded as follows:
\begin{equation}
    ||\tilde N_i|| \leq \rho(||z_i||)|| z_i||
    \label{equ:bound_Ni}
\end{equation}
where the outer-loop error signal $z_i \in \mathbb{R}^{12}$ is defined as
\begin{equation}
    z_i \triangleq \left[ \begin{matrix} e_{i1}^T & e_{i2}^T & r_i^T  & \tilde{\theta}_2 \end{matrix}\right]^T
\end{equation}
and $\rho: \mathbb{R}_{\geq 0}\rightarrow \mathbb{R}_{\geq 0}$ is a globally invertible, nondecreasing function. From Assumption 1, $||N_{\Delta_i}||$ and $||\dot N_{\Delta_i}||$ are bounded by positive constants:
\begin{equation}
    ||N_{\Delta_i}|| \leq \xi_{i}, \quad ||\dot N_{\Delta_i}|| \leq \dot \xi_i
\end{equation}

\noindent {\bf Lemma 1.} Let the auxiliary function $L_i(t) \in \mathbb{R}$ be defined as follows:
\begin{equation}
    L_i(t) \triangleq r_i^T\left(N_{\Delta_i} - \beta_i sgn(e_{i2})\right) + C_i
\end{equation}
If the control gain $\beta_i$ is selected to fulfill the following condition:
\begin{equation}
    \beta_i > \xi_i + \frac{1}{k_{i2}} \dot \xi_i
    \label{equ:betai_condition}
\end{equation}
 and $C_i \in \mathbb{R}^+$ is defined as 
 \begin{equation}
     C_i \triangleq  \sigma_1 \left( \frac{1}{\lambda_i} +\frac{1}{\varrho_2}\right)\xi_{\Delta_2}^2 + \sigma_2 \frac{1}{\varrho_2} \xi_{\Delta_2}
 \end{equation}
 where $\lambda_i < \varrho_2$ is a positive constant. Then $W_i \in \mathbb{R}$ defined by the following differential equation is always positive:
 \begin{equation}
    \begin{aligned}
    &\dot{W}_i \triangleq - \dot L_i \\
    & W_i(0) \triangleq \beta_i |e_{i2}(0)| - e_{i2}(0)N_{\Delta_i}(0)
    \end{aligned}
    \label{equ:Wi}
 \end{equation}
 The proof of Lemma 1 is similar to that given in \cite{8978475} and \cite{1310482}.
 
\noindent{\bf Theorem 1.} The inner-loop controller given in equation (\ref{equ:inner_input}), (\ref{equ:inner_RISE}), and (\ref{equ:update_law_inner}) ensures that signal $z_i$ is regulated that $||z_i(t)|| \rightarrow 0 $ as $t \rightarrow \infty$ provided that control gain $k_{s2}$ is selected sufficiently large, $k_{i1}, k_{i2} > \frac{1}{2}$, and $\beta_i$ following the condition (\ref{equ:betai_condition}).

{\it Proof.} Define an auxiliary vector $y \in \mathbb{R}^{13}$ as 
\begin{equation}
     y \triangleq \left[ \begin{matrix} z_i^T &\sqrt{W_i}\end{matrix} \right]^T
    \label{equ:y}
\end{equation}
and let $\mathcal{D} \subset \mathbb{R}^{13}$ be a domain containing $y(t) = \bf{0}$.
Define a Lyapunov function candidate as
\begin{equation}
    V_1(y,t) \triangleq \frac{1}{2} e_{i1}^Te_{i1} + \frac{1}{2} e_{i2}^Te_{i2} + \frac{1}{2}r_{i}^T J r_{i} + W_i + \frac{1}{2} \tilde{\theta}_i^T\Gamma^{-1} \tilde{\theta_2}
    \label{equ:V_1}
\end{equation}
where $V_1(y,t): \mathcal{D} \rightarrow \mathbb{R}$ is a positive definite, continuously differentiable function which satisfies 
\begin{equation}
    U_1(y)\leq V_1(y,t) \leq U_2(y)
    \label{equ:V1_bound}
\end{equation}
In equation (\ref{equ:V1_bound}), $U_1(y)$, $U_2(y) \in \mathbb{R}$ are continuous positive definite functions which are defined as
\begin{equation}
\begin{aligned}
& U_1(y) \triangleq c_1 ||y||^2 \\
& U_2(y) \triangleq c_2 ||y||^2 
\end{aligned}
\end{equation}
where $c_1$, $c_2 \in \mathbb{R}^+$ are defined as
\begin{equation}
\begin{aligned}
    & c_1 \triangleq \frac{1}{2}min \{1, \underline{J}, {\overline\Gamma}^{-1} \} \\
    & c_2 \triangleq \frac{1}{2}max \{1, \overline{J}, {\underline \Gamma}^{-1} \}
\end{aligned}
\label{equ:C1_2}
\end{equation}
In (\ref{equ:C1_2}), $\overline{J}$ and $\underline{J}$ indicate the maximum and minimum element of the diagonal matrix $J$ respectively.
The time derivative of $V_1(y,t)$ in (\ref{equ:V_1}) is expressed as
\begin{equation}
    \dot{V}_1 = e_{i1}^T \dot{e}_{i1} + e_{i2}^T \dot{e}_{i2} + r_i^T J \dot{r}_i + \dot{W}_i + \tilde{\theta}_2^T \Gamma^{-1} \dot{\tilde{\theta}}_2
    \label{equ:V1_d}
\end{equation}
Substituting equation (\ref{equ:inner_error}) and (\ref{equ:inner_close_error_d}) into the time derivative above, one has
\begin{equation}
    \begin{aligned}
    \dot{V}_1 = 
    & e_{i1}^T(e_{i2}-k_{i1}e_{i1}) + e_{i2}^T(r_i-k_{i2}e_{i2}) +\dot{W}_i + \tilde{\theta}_1^T \Gamma^{-1} \dot{\tilde{\theta}}_2 \\
    & + r_i^T(\tilde{N}_i(t) + N_{\Delta_i}-\dot{\mu}_2 - e_{i2}) 
    \end{aligned}
    \label{equ:V1_d1}
\end{equation}
With the definition of $\mu_2$ in (\ref{equ:inner_RISE_d}) and $W_i$ in (\ref{equ:Wi}), some of the terms in (\ref{equ:V1_d1}) can be eliminated, which yields
\begin{equation}
    \begin{aligned}
    \dot{V}_1 = 
    & -k_{i1}e_{i1}^2 - k_{i2}e_{i2}^2 + e_{i1}^T e_{i2} - (k_{s2} + 1)r_i^2 + r_i^T \tilde{N}_i(t) \\
    & - C_i + \tilde{\theta}_2^T \Gamma^{-1} \dot{\tilde{\theta}}_2
    \end{aligned}
    \label{equ:V1_d2}
\end{equation}
From equation (\ref{equ:update_law_inner}) and the definition of $\tilde{\theta}_2$, the time derivative of $\tilde{\theta}_2$ is expressed as
\begin{equation}
    \dot{\tilde{\theta}}_2  = \Gamma\left(\sigma_1 H_2 +\sigma_1 \frac{P_2^TH_2}{||P_2||} + \sigma_2 \frac{P_2^TH_2}{||P_2||\cdot ||H_2||}\right)
    \label{equ:d_est_error}
\end{equation}
Then (\ref{equ:V1_d2}) is expressed as
\begin{equation}
    \begin{aligned}
    \dot{V}_1 = 
    & -k_{i1}e_{i1}^2 - k_{i2}e_{i2}^2 + e_{i1}^T e_{i2} - (k_{s2} + 1)r_i^2 + r_i^T \tilde{N}_i(t) \\
    & - C_i + \tilde{\theta}_2^T \left(\sigma_1 H_2 +\sigma_1 \frac{P_2^TH_2}{||P_2||} + \sigma_2 \frac{P_2^TH_2}{||P_2||\cdot ||H_2||}\right) \\
    = & -k_{i1}e_{i1}^2 - k_{i2}e_{i2}^2 + e_{i1}^T e_{i2} - (k_{s2} + 1)r_i^2 + r_i^T \tilde{N}_i(t) \\ 
    & - C_i -\sigma_1\tilde{\theta}_2^T(P_2\tilde{\theta}_2 - \bar{\Delta}_2) - \sigma_1 \frac{(H_2 -\bar{\Delta}_2)^TH_2}{||P_2||}\\
    &- \sigma_2 \frac{(H_2 - \bar{\Delta}_2)^TH_2}{||P_2||\cdot ||H_2||}
    \end{aligned}
    \label{equ:V1_d3}
\end{equation}
By using Young's inequality and the bound of $\tilde{N}_i(t)$ in (\ref{equ:bound_Ni}),the following expressions are yielded
\begin{equation}
    \begin{aligned}
        & e_{i1}^T e_{i2} \leq \frac{1}{2}(||e_{i1}||^2 + ||e_{i2}||^2) \\
        & r_i^T \tilde{N}_i(t) \leq k_{s2}||r_i||^2 + \frac{1}{4k_{s2}}\rho^2(||z_i||)||z_i||^2 \\
        &-\sigma_1\tilde{\theta}_2^T(P_2\tilde{\theta}_2 - \bar{\Delta}_2) - \sigma_1 \frac{(H_2 -\bar{\Delta}_2)^TH_2}{||P_2||}-\sigma_2 \frac{(H_2-\bar{\Delta}_2)^TH_2}{||P_2||\cdot ||H_2||} \\
        & \leq -\sigma_1( \varrho_2- \lambda_i)||\tilde{\theta}_2||^2 + C_i
    \end{aligned}
    \label{equ:Young}
\end{equation}
Substituting (\ref{equ:Young}), (\ref{equ:V1_d3}) is upper bounded as
\begin{equation}
\begin{aligned}
    \dot{V}_1 \leq 
    & - (k_{i1} - \frac{1}{2}) ||e_{i1}||^2 - (k_{i2} - \frac{1}{2}) ||e_{i2}||^2 - ||r_i||^2 \\
    & + \frac{1}{4k_{s2}}\rho^2(||z_i||)||z_i||^2 - \sigma_1( \varrho_2- \lambda_i)||\tilde{\theta}_2||^2 \\
    \leq & -\left(c_3 - \frac{1}{4k_{s2}}\rho^2(||z_i||)\right)||z_i||^2
\end{aligned}
    \label{equ:V1_bound1}
\end{equation}
where $c_3 \in \mathbb{R}$ is defined as
\begin{equation}
    c_3 \triangleq min \left\{k_{i1} - \frac{1}{2}, k_{i2} - \frac{1}{2},1,\sigma_1(\varrho_2- \lambda_i)\right\}
\end{equation}
$c_3$ is positive provided the definition of $\lambda_i$ in {\it Lemma 1.} The expression in (\ref{equ:V1_bound1}) can be further upper bounded as
\begin{equation}
    \dot{V}_1 \leq -c_i ||y||^2, \quad \forall y \in \mathcal{D}_1 
\end{equation}
for some positive constant $c_i$. Set $\mathcal{D}_1 \subset \mathcal{D}$ is defined as 
\begin{equation}
    \mathcal{D}_1 \triangleq \left\{y(t) \in \mathbb{R}^{13} \mid ||y(t)|| \leq \rho^{-1}\left(2\sqrt{c_3k_{s2}}\right) \right \}
    \label{equ:V1_bound2}
\end{equation}
The inequality (\ref{equ:V1_bound2}) shows that $V_1(y,t) \in \mathcal{L}_{\infty}$ in $\mathcal{D}_1$; hence  $e_{i1}$, $e_{i2}$, $r_i$, and $\tilde{\theta}_2\in \mathcal{L}_{\infty}$ in $\mathcal{D}_1$. Similar to proof in \cite{6579806}, The attraction region $\mathcal{R}_1 \subset \mathcal{D}_1$ as 
\begin{equation}
    \mathcal{R}_1 \triangleq \left\{y(t) \in \mathbb{R}^{13} \mid U_2(y) \leq c_1 \left(\rho^{-1}\left(2\sqrt{c_3k_{s2}}\right)\right)^2 \right \}
\end{equation}
Hence, $||y(t)|| \rightarrow 0$ as $t\rightarrow \infty, \forall y(0) \in \mathcal{R}_1$, which further indicates that $||z_i(t)|| \rightarrow 0$ as $t\rightarrow \infty, \forall y(0) \in \mathcal{R}_1$.

\noindent{\bf Corollary 1.} $P_2$ is upper bounded by $||P_2|| \leq \xi_{p2}$, where $\xi_{p2} \in \mathbb{R}^+$ is a positive constant. 

{\it Proof.} From Theorem 1, $||e_{i1}|| \rightarrow 0$ as $t \rightarrow 0$. Since $||\Psi_{2d}||$ is bounded in Property 1, the continuous function $\Psi_{2f}$ is upper bounded by $||\Psi_{2f}|| \leq \zeta_2$, where $\xi_2 \triangleq \frac{1}{l_2}\zeta_2^2 + ||\varrho_2||\in \mathbb{R}^+$ is a positive constant. From equation (\ref{equ:PQ_inner}),
\begin{equation}
    ||P_2|| = e^{-l_2t}\left\|\int_0^t e^{l_2\tau)}\Psi_{2f}^T(\tau) \Psi_{2f}(\tau) d \tau \right\|  + \left\|\varrho_2 E_3\right\|
\end{equation}
The norm of $P_2$ is upper bounded by
\begin{equation}
\begin{aligned}
    ||P_2|| & \leq e^{-l_2 t}\zeta_2^2 \int_0^te^{l_2 \tau}d\tau + \varrho_2 \\
    & \leq \frac{1}{l_2}\zeta_2^2 + \varrho_2
\end{aligned}
\end{equation}
The corollary is proved.

\noindent{\bf Theorem 2.} For error system (\ref{equ:inner_close_error_d}) with the adaptive estimation law given in (\ref{equ:update_law_inner}), the estimation error variable $P_2^{-1}H_2$ is regulated that $||P_2^{-1}H_2|| \rightarrow 0 $ in a finite time $t_1$ if $\Gamma$, $\sigma_1$, $\sigma_2$ and $\varrho_2$ are properly selected (see the subsequent proof). And the estimation error $\tilde{\theta}_2$ is guaranteed to converge to a compact set around zero in $t_i$.

{\it Proof.} Let $\Xi \subset \mathbb{R}^3$ be a domain containing $||P_2^{-1}(t)H_2(t)|| = 0$. Define a Lyaponov candidate as
\begin{equation}
    V_2(P_2^{-1}H_2) \triangleq \frac{1}{2}H_2^T P_2^{-1}P_2^{-1}H_2
    \label{equ:V2}
\end{equation}
where $V_2(P_2^{-1}H_2): \Xi \rightarrow \mathbb{R}_{\geq 0}$ is a positive definite, continuously differentiable function satisfies a similar condition to (\ref{equ:V1_bound}).
The time derivative of $V_2$ is expressed as
\begin{equation}
    \dot{V}_2 = H_2^T P_2^{-1} \frac{\partial}{\partial t} \left(P_2^{-1} H_2\right)
    \label{equ:V2_d}
\end{equation}
From equation (\ref{equ:H2_Delta}), one has
\begin{equation}
    P_2^{-1}H_2 = -\tilde{\theta}_2 + P_2^{-1}\bar{\Delta}_2
    \label{equ:PH}
\end{equation}
Exploiting $\frac{\partial}{\partial t}P_2^{-1} = -P_2^{-1}\dot{P}_2 P_2^{-1}$, and substituting equation (\ref{equ:d_est_error}) and (\ref{equ:PH}) into (\ref{equ:V2_d}), $\dot{V}_2$ can be expressed as
\begin{equation}
\begin{aligned}
    \dot{V}_2 = 
    & - H_2^TP_2^{-1}\left(\Gamma \sigma_1 H_2 + \Gamma \sigma_1 \frac{P_2^T H_2}{||P_2||}\right) \\
    &- H_2^TP_2^{-1}\left( \Gamma \sigma_2 \frac{P_2^T H_2}{||P_2||\cdot ||H_2||} - \Phi \right)
\end{aligned}
    \label{equ:V2d_2}
\end{equation}
where $\Phi \in \mathbb{R}^3$ is defined as
\begin{equation}
    \Phi \triangleq - P_2^{-1}\dot{P}_2P_2^{-1} \bar{\Delta}_2 + P_2^{-1} \dot{\bar{\Delta}}_2
\end{equation}
From  Assumption 2, Property 4, and Corollary 1, $\Phi$ is verified to be bounded, and $\dot{V}_2$ is upper bounded as
\begin{equation}
\begin{aligned}
    \dot{V}_2 \leq &-\left(\underline{\Gamma} \sigma_2 \frac{1}{\xi_{p2}} - \frac{1}{\varrho_2}\left\|\Phi\right\|\right)\left\|H_2\right\| - \frac{2}{\xi_{p2}}\underline{\Gamma} \sigma_1 \left\|H_2\right\|^2 \\
    \leq & -\sqrt{2}\varrho_2\left(\underline{\Gamma} \sigma_2 \frac{1}{\xi_{p2}} - \frac{1}{\varrho_2}\left\|\Phi\right\|\right)\sqrt{V_2} - 4\frac{\varrho_2^2}{\xi_{p2}}\underline{\Gamma} \sigma_1 V_2
\end{aligned}
\label{equ:V2_d3}
\end{equation}
If $\sigma_2$ is selected sufficiently large and $\Gamma$, $\varrho_2$ are selected to satisfy 
\begin{equation}
    \underline{\Gamma} \sigma_2 \frac{1}{\xi_{p2}} - \frac{1}{\varrho_2}\left\|\Phi\right\| > 0
    \label{equ:gamma_cri}
\end{equation}
the expression in (\ref{equ:V2_d3}) can be further upper bounded as
\begin{equation}
    \dot{V}_2 \leq - c_{i1} \sqrt{V_2}-c_{i2} V_2, \quad \forall P_2^{-1}H_2 \in \Xi_1
\end{equation}
where $c_{i1}$, $c_{i2} \in \mathbb{R}^+$ are positive constants, and $\Xi_1$ can be made arbitrarily large by increasing $\sigma_2$ and select $\Gamma$, $\varrho_2$ based on the design criteria in (\ref{equ:gamma_cri}). Similar to the proof of Theorem 1, an attraction region $\mathcal{R}_{\Xi} \subset \Xi_1$ exists that $||P_2^{-1}(t)H_2(t)|| \rightarrow 0$ in $t_i \leq \frac{2}{c_{i2}} ln\left(1 + \frac{c_{i2}}{c_{i1}}a_i\right)$, $\forall P_2^{-1}(0)H_2(0) \in \mathcal{R}_{\Xi}$ where $a_i \in \mathbb{R}^+$ is defined as $a_i     \triangleq \frac{\sqrt{2}}{2}\left(||\tilde{\theta}_2(0)|| + \frac{1}{\varrho_2}\xi_{\Delta_2}\right)$.

From the definition of $H_2$ in equation(\ref{equ:H2_Delta}), this further implies that 
$\tilde{\theta}_2$ converges to a compact set $\mathcal{R}_i$ in $t_i$, where $\mathcal{R}_i \subset \mathbb{R}^3$ is defined as
\begin{equation}
    \mathcal{R}_i \triangleq \left\{\tilde{\theta}_2(t) \in \mathbb{R}^3 \mid ||\tilde{\theta}_2|| \leq \frac{1}{\varrho_2}\xi_{\Delta_2}\right\}
\end{equation}
This completes the proof.

Notice that though the FT convergence can be guaranteed by properly selecting $\Gamma$, $\sigma_2$, and $\varrho_2$ while $\sigma_1$ is selceted as zero, the selection of learning gain $\sigma_1$ also influences the convergence rate of $||P_2^{-1}H_2||$. A large $\sigma_1$ leads to a faster convergence. However, it might also cause oscillations in the estimated parameters if $\sigma_1$ is designed too large.

\subsection{Outer-Loop Analysis}
The time derivative of equation (\ref{equ:outer_close_error_dynamics}) is calculated similar to (\ref{equ:inner_close_error_d}) as
\begin{equation}
    J\dot{r}_o = \tilde{N}_o +N_{\Delta_1} - \dot{\mu}_1 - e_{o2}
    \label{equ:outer_close_error_d}
\end{equation}
where auxiliary functions $\tilde{N}_o$, $N_{\Delta_1}$ $\in \mathbb{R}^3$ are defined similar to (\ref{equ:Ni}). The inner loop error signal $z_o \in \mathbb{R}^{10}$ is defined as
\begin{equation}
    z_o \triangleq \left[ \begin{matrix} e_{o1}^T & e_{o2}^T & r_o^T  & \tilde{\theta}_1 \end{matrix}\right]^T
\end{equation}
The subsequent theorems can be proved.

\noindent{\bf Theorem 3.} The outer-loop controller given in equation (\ref{equ:outer_input}), (\ref{equ:outer_RISE}), and (\ref{equ:update_law_outer}) ensures that signal $z_o$ is regulated that $||z_o(t)|| \rightarrow 0 $ as $t \rightarrow \infty$ provided that control gain $k_{s1}$ is selected sufficiently large, $k_{o1}, k_{o2} > \frac{1}{2}$, and $\beta_o$ following a condition similar to (\ref{equ:betai_condition}).

\noindent{\bf Theorem 4.} For error system (\ref{equ:outer_close_error_d}) with the adaptive estimation law given in (\ref{equ:update_law_outer}), the estimation error variable $P_1^{-1}H_1$ is regulated that $||P_1^{-1}H_1|| \rightarrow 0 $ in a finite time $t_o$ satisfying $t_o \leq \frac{2}{c_{o2}} ln\left(1 + \frac{c_{o2}}{c_{o1}}a_o\right)$ if $\gamma$, $\gamma_1$, and $\varrho_1$ are properly selected, where $a_o \in \mathbb{R}$ is defined as $a_o \triangleq \frac{\sqrt{2}}{2}\left(\tilde{\theta}_1(0) + \frac{1}{\varrho_1}\xi_{\Delta_1}\right)$. And $\tilde{\theta}_1$ is guaranteed to converge to a compact set around zero in $t_o$.

{\it Proof of Theorem 3 and 4.} Noticing that $\mathrm{sat}\left(H_1\right) \leq 1$, the proof can be conducted in a similar method as that of the inner-loop controller. 

\section{SIMULATION}
In this section, the effectiveness of the designed controller is verified by simulations and experiments. Comparative simulations are carried out between the proposed strategy and the traditional methods based on SMC and gradient algorithms in \cite{7152250}. The results show that the proposed method yields better tracking error and estimation convergence. Meanwhile, it generates smoother input signals for practical applications than SMC controllers. The results indicate the robustness against disturbances and mass-inertia changes of the controller.

To verify the performance of the proposed control strategy, numrical simulations are conducted in MATLAB. Table (\ref{tab:param}) shows the preset mass-inertia parameters of the UAV in simulation.
\begin{table}[htbp]
    \centering
    \begin{tabular}{c c c}
    \toprule
        Item & Quantity & Unit \\
    \midrule
        m & 3.12 & kg \\
        $I_{x}$ & 0.1 & $kg \cdot m^2$ \\
        $I_{y}$ & 0.1 & $kg \cdot m^2$ \\
        $I_{z}$ & 0.2 & $kg \cdot m^2$ \\
    \bottomrule
    \end{tabular}
    \caption{True value of mass-inertia parameters}
    \label{tab:param}
\end{table}
The control gains and learning gains of the proposed RISE-based adaptive controller with mass-inertia estimation (RISE-Emi) is selected as the following table (\ref{tab:gains}). And the learning gain matrix $\Gamma$ is selected as
\begin{equation}
    \Gamma =
    \left[ \begin{matrix} 10^{-4} & 0 & 0 \\ 0 & 10^{-4} & 0 \\ 0 & 0 & 4.5\times 10^{-3} \\\end{matrix} \right]
\end{equation}

\begin{table}[htbp]
    \centering
\begin{tabular}{cc|cc}
\toprule
\multicolumn{2}{c|}{Outer-loop gains} & \multicolumn{2}{c}{Inner-loop gains} \\
\midrule
Symbol               & Value         & Symbol               & Value         \\
\midrule
\centering
$k_{o1}$        &   1      & $k_{i1}$        &    2       \\
$k_{o2}$        &   1      & $k_{i2}$        &    1       \\
$k_{s1}$        &   5.4    & $k_{s2}$        &    4.5     \\
$\beta_o$       &   1      & $\beta_i$       &    1       \\
$\alpha_1$      &   3      & $\alpha_2$      &    5       \\
$\varrho_1$     &   0.5    & $\varrho_2$     &    0.5     \\
$\gamma$        &   0.3    & $\sigma_1$      &    8       \\
$\gamma_1$      &   0.17   & $\sigma_2$      &    200     \\
\bottomrule
\end{tabular}
\caption{table: Control and learning gains}
\label{tab:gains}
\end{table}

The comparison is conducted between RISE-Emi and the adaptive sliding mode controller with gradient-based mass estimation (ASMC) proposed in \cite{7152250}. We selcet the desired trajectry and yaw angle as
\begin{equation}
\begin{aligned}
    &\eta_{1d} = 2sin(t) \cdot \left[ \begin{matrix} 1 & 1 & 1\end{matrix}\right]^T \\
    &\psi_d = sin(1.1t)
\end{aligned}
\end{equation}
and add white noise disturbance to the dynamic model output of the system.

\begin{figure}[htbp]
    \centering
    \includegraphics[width=1\linewidth]{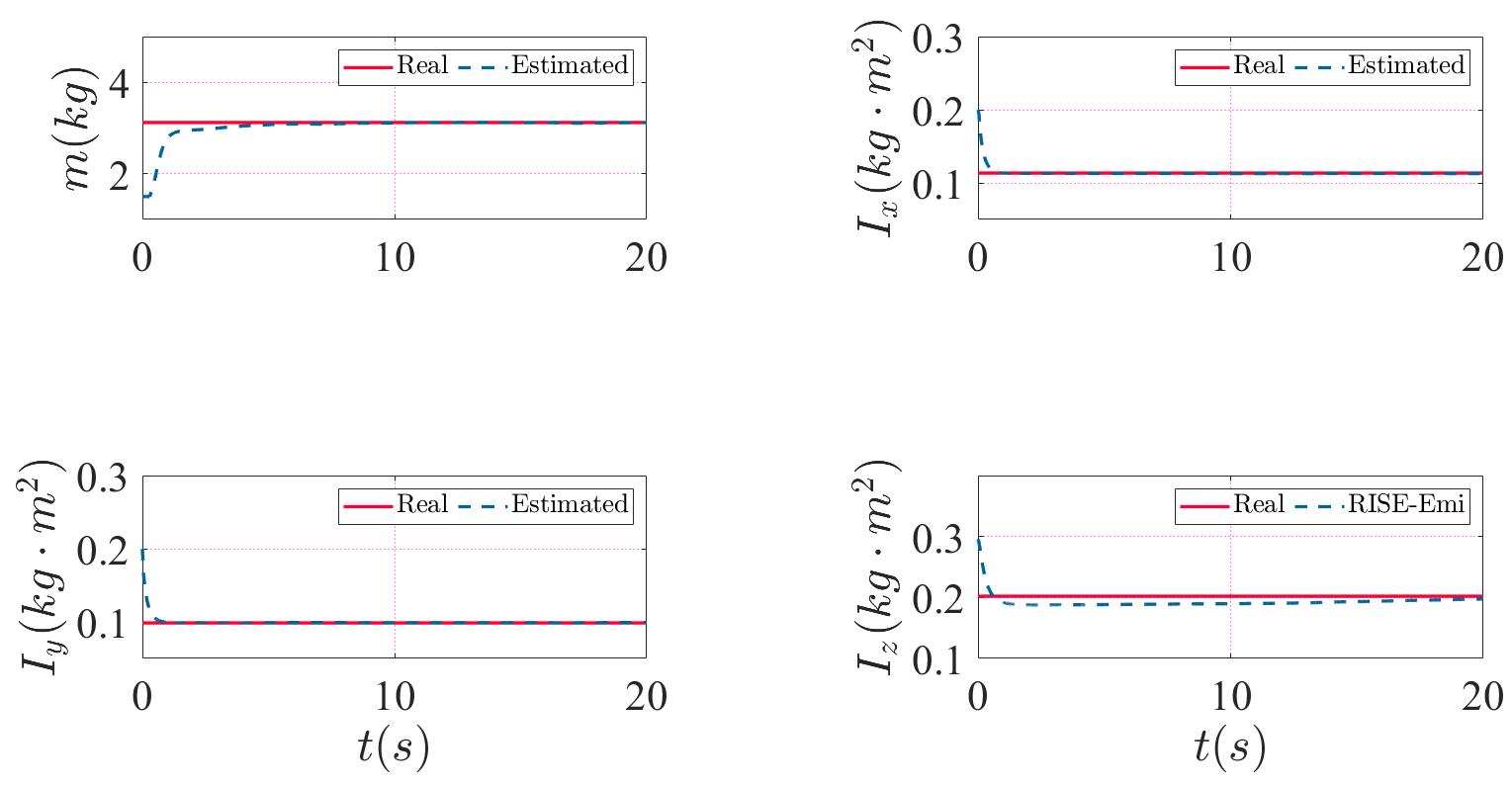}
    \caption{Parameter estimation results}
    \label{fig:est_my}
\end{figure}

\begin{figure}[htbp]
    \centering
    \includegraphics[width=\linewidth]{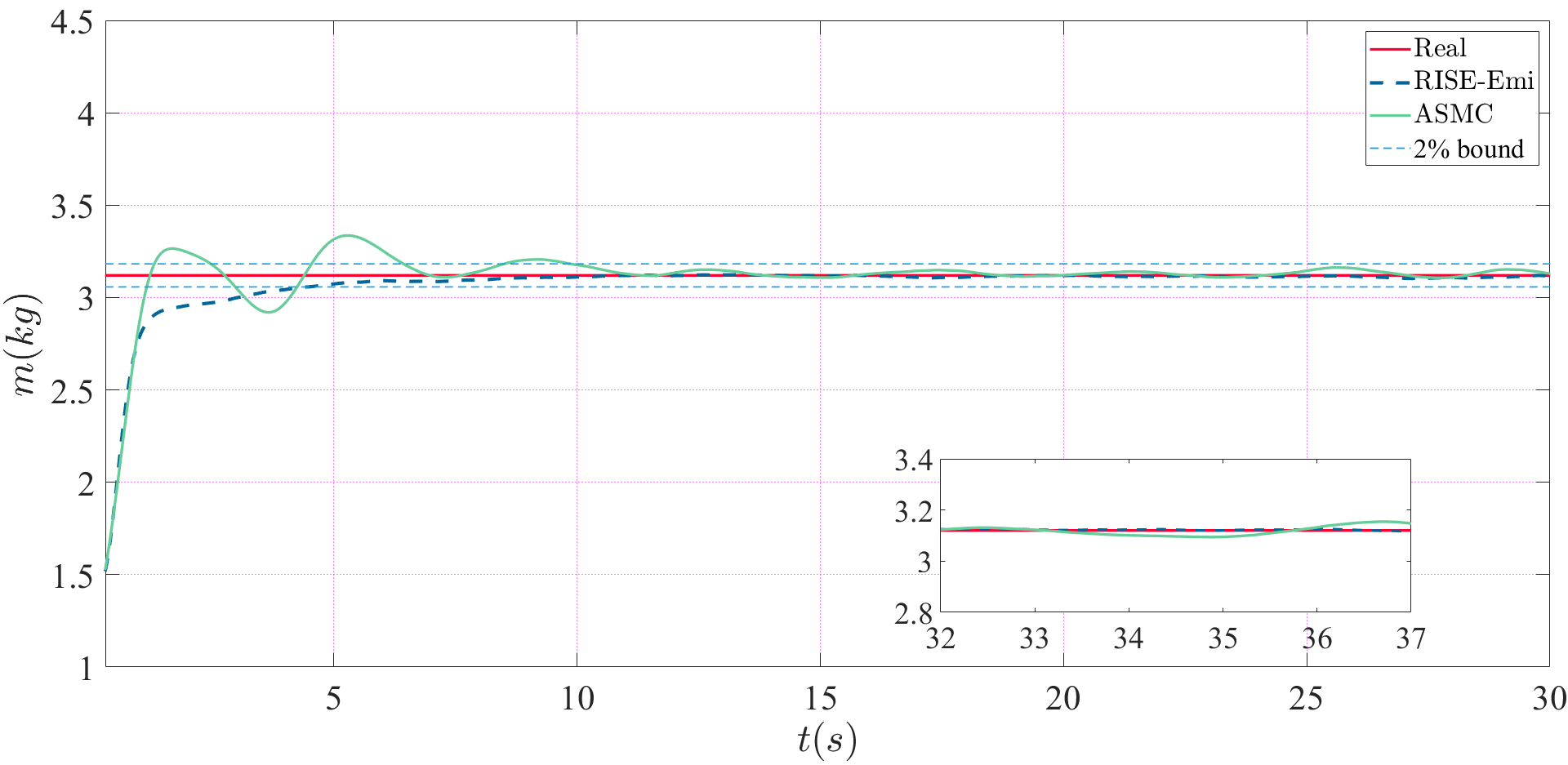}
    \caption{Comparison of estimation of mass}
    \label{fig:est_comp}
\end{figure}

The result of mass-inertia estimation of RISE-Emi is show in Fig. \ref{fig:est_my}. The initial estimation of mass is set $50\%$ smaller than the real value, and the initial inertia estimations are about $100\%$, $100\%$ and $50\%$ larger respectively. All 4 estimatied values converge to its truth finally. Due to the different dynamic characters between yaw orientation and roll, pitch orientation, the estimation of $I_z$ overshoots for about $5\%$ with the selected parameters. And the convergence of mass is relatively slow because of the slower response of the outer loop compared to the inner loop.
Meanwhile, the estimation error is compared with the mass estimation of ASMC in Fig. \ref{fig:est_comp}. Initially, the estimated mass of the 2 methods converge at a similar speed. Then, the estimation in RISE-Emi achieves the $2\%$ bound faster, and gradually reaches the real value within about $10s$, while keeps increasing at a large speed and saturates before convergence in ASME. It is also noticable that because of the added white noise, the steady-state error can not be zero all the time. However, the error caused by the noise in RISE-Emi is smaller than that of ASMC, which is shown in the sub-figure of Fig \ref{fig:est_comp}.

These results indicate the effectiveness of our method in mass-inertia estimation and its robustness against disturbances.

\begin{figure}[htbp]
    \centering
    \includegraphics[width=\linewidth]{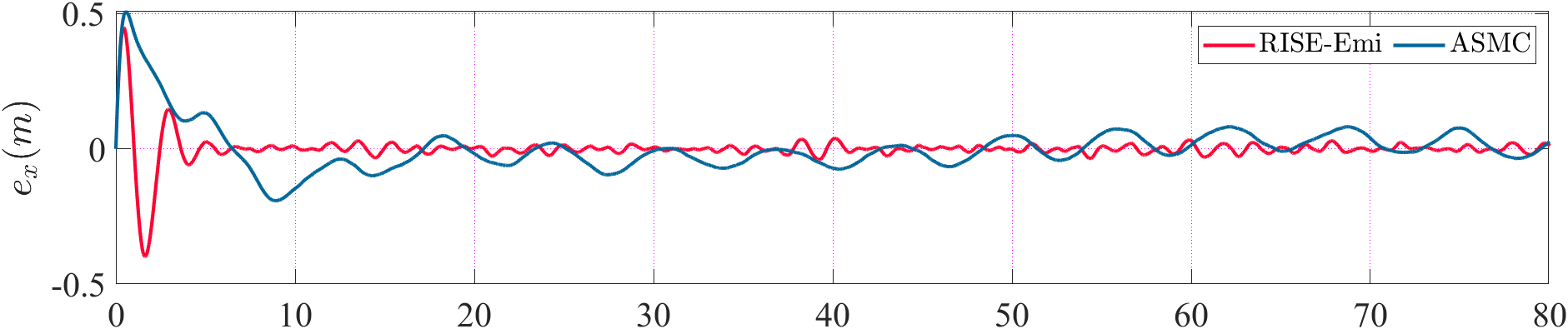}
    \includegraphics[width=\linewidth]{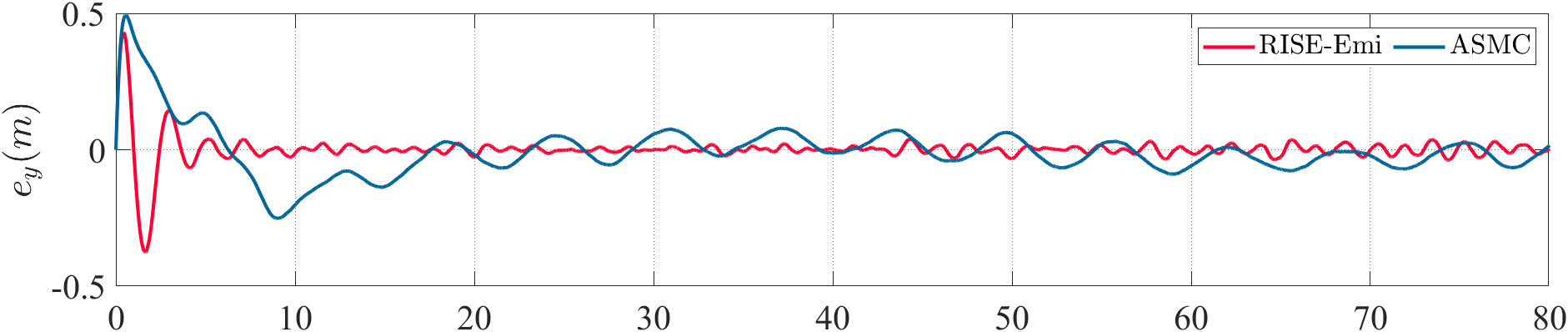}
    \includegraphics[width=\linewidth]{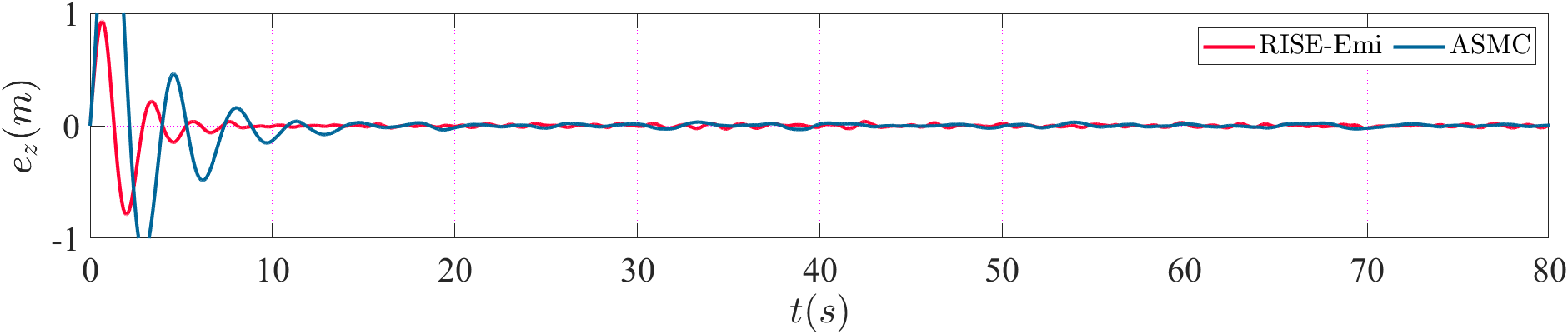}
    \caption{Position tracking errors}
    \label{fig:pe}
\end{figure}

\begin{figure}[htbp]
    \centering
    \includegraphics[width=\linewidth]{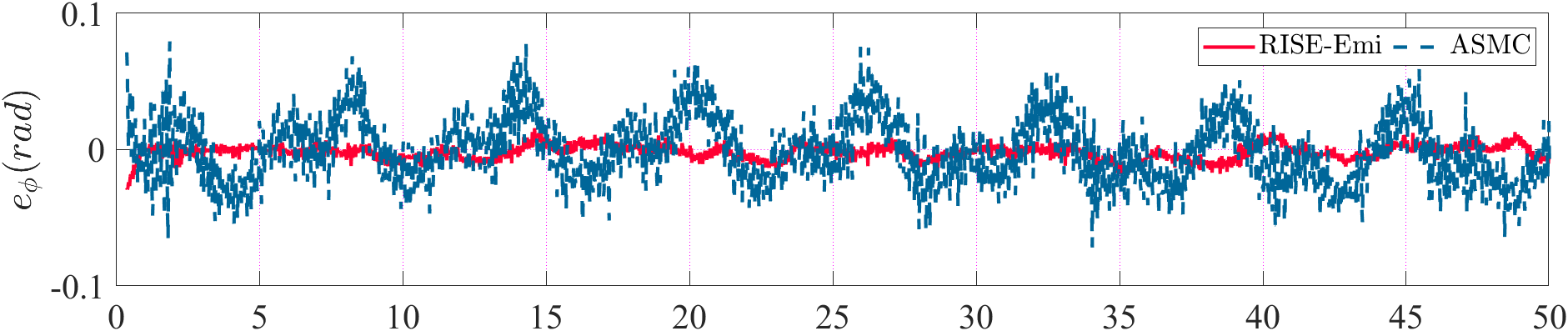}
    \includegraphics[width=\linewidth]{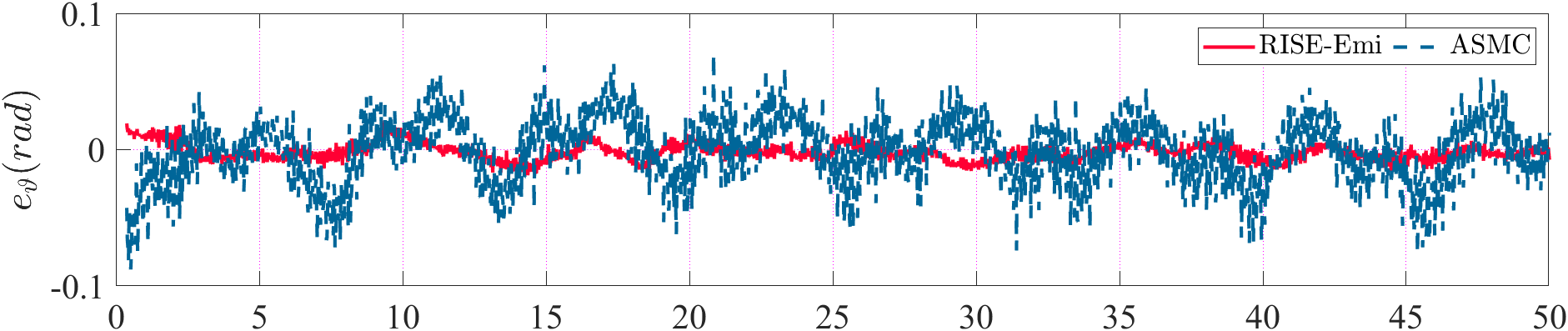}
    \includegraphics[width=\linewidth]{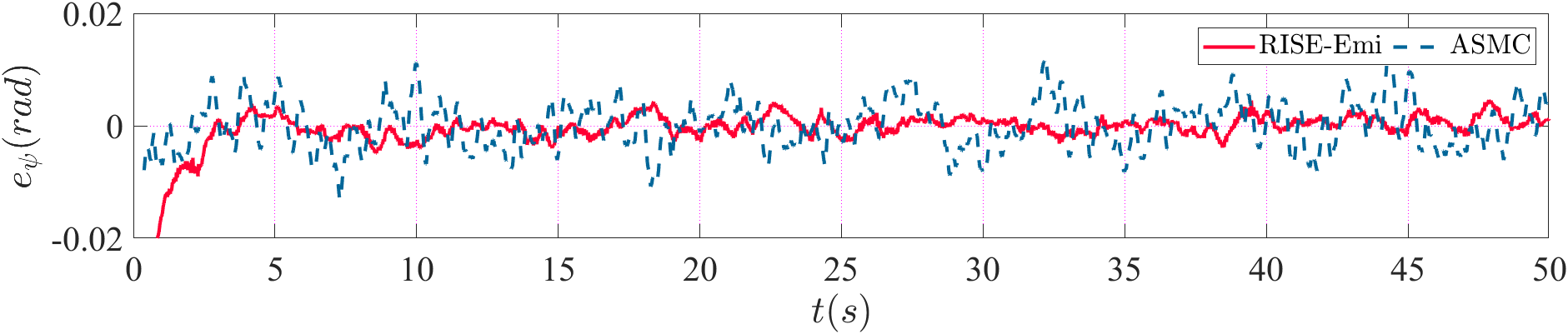}
    \caption{Attitude tracking errors}
    \label{fig:ae}
\end{figure}

Then, the comparison of trajectory and attitude tracking errors are provided in Fig. \ref{fig:pe}. and Fig. \ref{fig:ae} respectively. The trajctory tracking errors increase greatly in the beginning mainly because of the imprecise initial estimation, which undermines the performance of the controllers. Then, the tracking errors of the proposed method converge faster than ASMC, also yielding smaller steady-state errors in $x$ and $y$ directions. And it is obvious that the scale of attitude tracking error in RISE-Emi is much smaller, which illustrate the disturbance rejection ability of our method.

\begin{figure}[htbp]
    \centering
    \includegraphics[width=\linewidth]{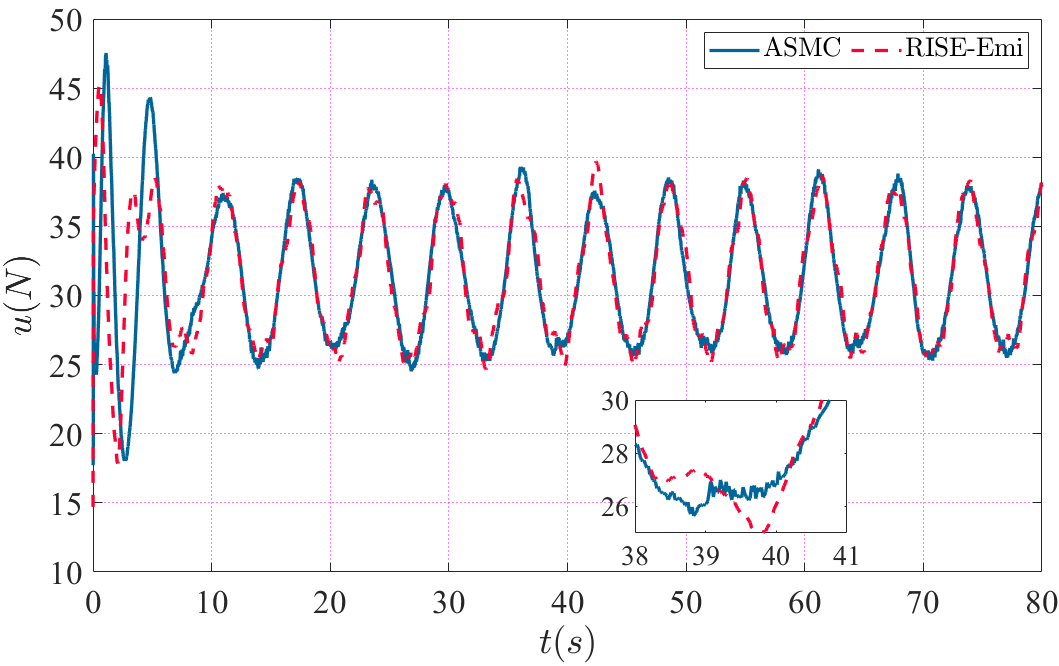}
    \caption{Comparison of thrust}
    \label{fig:thrust_comp}
\end{figure}

Finally, the thurst output of the controller is compared in Fig. \ref{fig:thrust_comp}. The output of the proposed method is smoother than that of ASMC when disturbance exists, which is more physically achievable in practical applications.
\section{CONCLUSION}
In this work, we have developed and validated an adaptive control strategy for UAVs in face of external disturbances and mass-inertia variation. First, a dynamic model of multi-rotor UAVs with disturbances is derived with a linearly parameterized form. Then, a cascade control law is designed based on this form with robust RISE terms. Finally, mass-inertia estimation is conducted based on a filtering operation to improve the robustness against possible mass-inertia change. Comparative simulations have shown that a better performance can be achieved with our method than the previously proposed method ASMC.

\section*{ACKNOWLEDGEMENT}
This work is motivated by Haoxuan Shan's work of a integrated quadruped-hexarotor system. Gang Chen has also contributed to the idea and process of the research.

\bibliographystyle{./IEEEtran} 
\bibliography{./IEEEabrv,./MyRef}
\end{document}